\documentclass[%
reprint,
nofootinbib,
 amsmath,amssymb,
 aps,
]{revtex4-1}
\usepackage{epsfig,epsf,rotating,wrapfig,tabularx}
\usepackage{dcolumn}
\usepackage{bm}
\usepackage{graphicx}

\begin{document}

\preprint{APS/123-QED}

\title{A variation of the charged particle spectrum shape as function of rapidity in high energy $pp$ collisions}

\author{\firstname{A.~A.}~\surname{Bylinkin}}
 \email{alexander.bylinkin@desy.de}
\affiliation{%
 Institute for Theoretical and Experimental
Physics, ITEP, Moscow, Russia
}%
\author{\firstname{A.~A.}~\surname{Rostovtsev}}
 \email{rostov@itep.ru}
\affiliation{%
 Institute for Theoretical and Experimental
Physics, ITEP, Moscow, Russia
}%


\begin{abstract}
The shapes of invariant differential cross section for charged particle production as function of transverse momentum measured in $pp$ collisions by the UA1 detector are analyzed. The spectra shape varies with the produced particle's pseudorapidity changing. To describe this and several other recently observed effects a simple qualitative model for hadroproduction mechanism was proposed.
\end{abstract}

\pacs{Valid PACS appear here}
\maketitle



\begin{figure}[h]
\includegraphics[width =8cm]{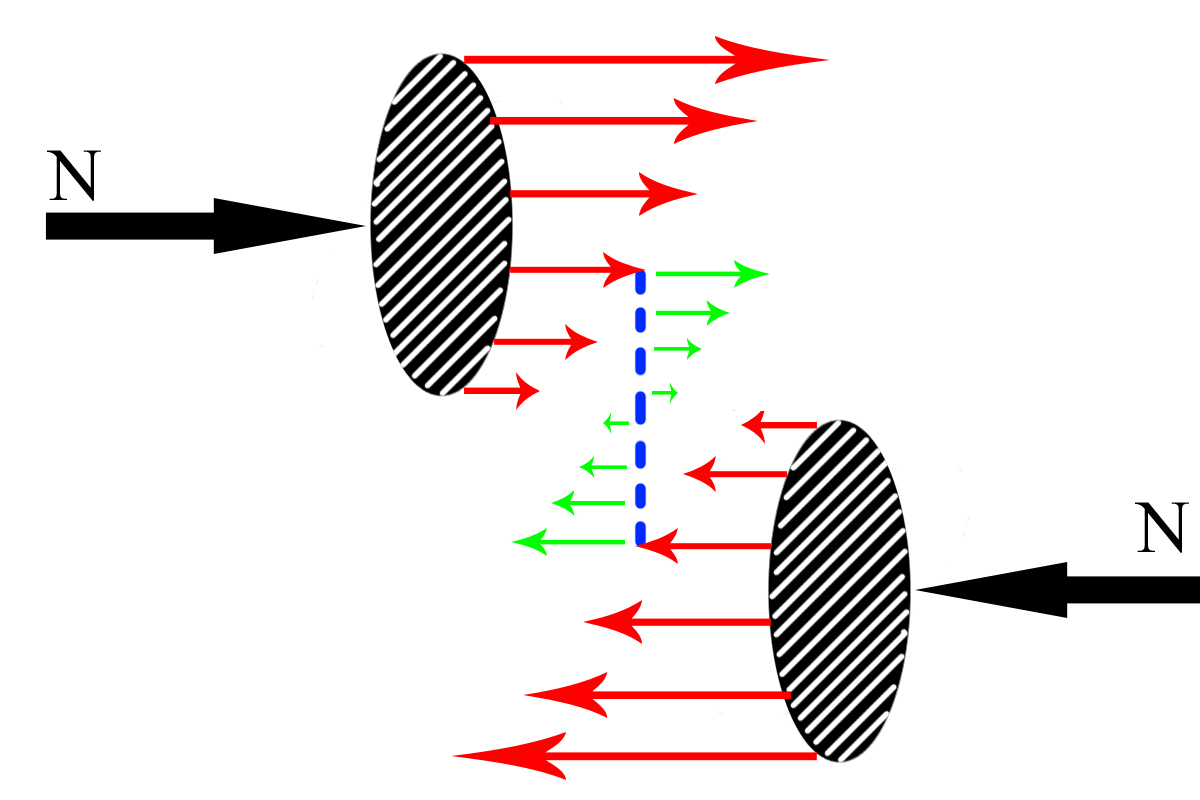}
\caption{\label{fig} Two different sources of hadroproduction: red arrows - particles produced by the preexisted partons, green - particles produced via the Pomeron exchange.}
\end{figure}

The baryon-baryon high energy interactions one could decompose into at least two distinct sources of produced hadrons. The first one is associated with the baryon valence quarks and a quark-gluon cloud coupled to the valence quarks. Those partons preexist long time before the interaction and could be considered as being a thermalized statistical ensemble. When a coherence of these partonic systems is destroyed via strong interaction between the two colliding baryons these partons hadronize into particles released from the collision. The hadrons from this source are distributed presumably according to the Boltzmann-like exponential statistical distribution in transverse plane w.r.t. the interaction axis. The second source of hadrons is directly related to the virtual partons exchanged between two colliding partonic systems. In QCD this mechanism is described by the BFKL Pomeron exchange. The radiated partons from this Pomeron have presumably a typical for the pQCD power-law spectrum. Schematically Figure~\ref{fig}  shows these two sources of particles produced in high energy baryonic collisions. 

Recently, a new unified approach to describe the particle production spectra shape was proposed~\cite{OUR1}. It was suggested to approximate the charged particle spectra as function of the particle’s transverse momentum~($P_T$) by a sum of an exponential (Boltzmann-like) and a power law statistical distributions. 
\begin{equation}
\label{eq:exppl}
\frac{d\sigma}{P_T d P_T} = A_e\exp {(-E_{Tkin}/T_e)} +
\frac{A}{(1+\frac{P_T^2}{T^{2}\cdot n})^n},
\end{equation}
where  $E_{Tkin} = \sqrt{P_T^2 + M^2} - M$
with M equal to the produced hadron mass. $A_e, A, T_e, T, n$ are the free parameters to be determined by fit to the data.  The detailed arguments for this particular choice are given in~\cite{OUR1}.  

\begin{figure}[h]
\includegraphics[width =8cm]{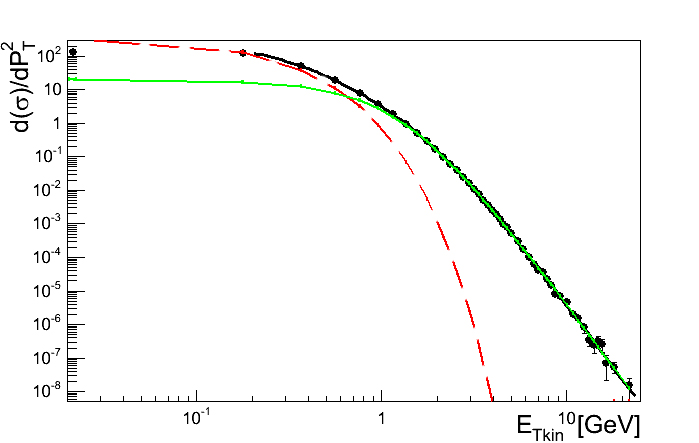}
\caption{\label{fig:0} Charge particle differential cross section~\cite{UA1} fitted to the function~(\ref{eq:exppl}): the red (dashed) line shows the exponential term and the green (solid) one - the power law.}
\end{figure}

The proposed new parameterization matches the naive illustrative picture of hadroproduction in baryon-baryon collisions described above.
A typical charged particle spectrum as function of transverse energy, fitted with this function~(\ref{eq:exppl}) is shown in Fig~\ref{fig:0}. 
As one can see the exponential term dominates the particle spectrum at low $P_T$ values.

The contributions of the exponential and power law
terms
of the parameterization~(\ref{eq:exppl}) to
the typical spectrum of charged particles produced in $pp-$collisions are also shown separately in Figure~\ref{fig:0}.
The relative contribution of these terms is characterized by ratio $R$ of the power law term alone to the parameterization function integrated over $P_T^{2}$:

\begin{equation}
R = \frac{AnT}{AnT + A_e(2MT_e + 2T_e^2)(n-1)}
\end{equation}

It was found that the exponential contribution dominates the charged particle spectra in baryon-baryon collisions while it is completely missing in gamma-gamma interactions~\cite{OUR1}. Moreover, the exponential contribution is characteristic for charged pion production and is much less pronounced in kaon or proton (antiproton) production spectra~\cite{OUR2}. There is also no room for the exponential contribution in a heavy quarkonium production in pp collisions~\cite{CDF}. In addition, the power law term contribution increases with the event charged multiplicity and energy of baryon-baryon collisions~\cite{OUR4}. These observations are naturally fit into the simple hadroproduction picture described above.


Indeed:

 1. In gamma-gamma interactions there are no preexisted partons to form the exponential part of the produced particle spectra.

2.	In $pp$-collisions the BFKL Pomerons are more flavor democratic with respect to the valence quark related radiation. This results in much smaller exponential contribution to the charged kaon spectra produced in baryon-baryon interactions then that to the charged pion spectra. Such behavior recently proved to be true by the comparative analysis~\cite{OUR3} of the data provided by the PHENIX experiment for $pp$ collisions at RHIC~\cite{Adare:2011vy}.  Moreover, it turned out that the difference in the relative contribution of the exponential part in the approximation (\ref{eq:exppl}) for pions and kaons defines the peculiar shape of the kaon to pion yield ratio as function of the hadron transverse momentum.   

3.	The AGK cutting rules~\cite{AGK} state that charge multiplicity in hadronic interactions is proportional  to the number of Pomerons involved in this interaction. Therefore, the relative contribution of the exponential part of the approximation (\ref{eq:exppl}) will decrease with the increase of charged multiplicity in $pp$ interactions. This has been recently checked  in the analysis~\cite{OUR4} of charged particle spectra for the $pp$-collision events with different visible charged multiplicity published  by the STAR experiment~\cite{STAR}.

    Moreover, one could make further predictions about some observable features of the hadroproduction phenomenon:

1.	In high energy photon-proton collision the exponential contribution to the charged particle spectra will strongly depend on rapidity of produced charged particles. The hadrons produced in the proton direction in rapidity space will show a sizable exponential contribution to their spectra, while the distribution of hadrons produced on the photon side of the event will be described by the power law function only. Therefore, one expects to observe a change between these two regimes for particles produced around zero rapidities in the photon-proton center of mass system. 

2.	Another prediction could be made about an absolute dominance of the exponential contribution to the spectra of particles produced in the high rapidity proton fragmentation region where the role of valence quark is more important.

The prediction (1) could be checked with a detailed measurement of the photon-proton interaction at HERA experiments. For that the double differential distribution of produced charged particles in transverse momentum and rapidity space has to be measured in the photon-proton center of mass system. 

\begin{figure}[h]
\includegraphics[width =8cm]{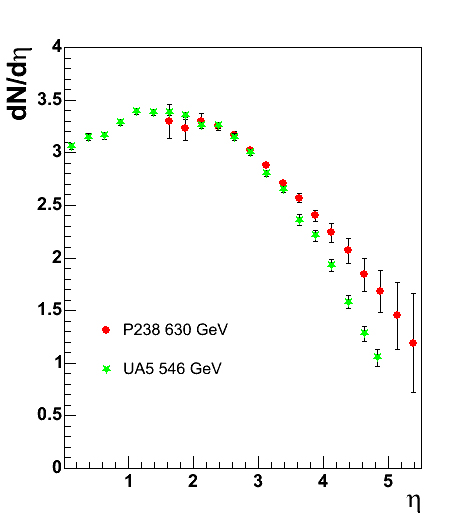}
\caption{\label{fig:01} Pseudorapidity distribution of charged particles in $p\bar{p}$ collisions at $\sqrt{s} = 630 GeV$~\cite{P238} and $\sqrt{s} = 546 GeV$~\cite{UA5}.}
\end{figure}

To check the prediction (2)  it is possible to use already available data published by the UA1 experiment~\cite{UA1}. These data are presented by charged particle  spectra for $pp$-collision in 5 different pseudorapidity regions, covering the total rapidity interval $|\eta|<3.0$. Pseudorapidity distributions of charged particles in $p\bar{p}$ collisions~\cite{P238, UA5} with the center of mass energy close to that in UA1 experiment~\cite{UA1} are shown in figure \ref{fig:01}. The rapidity central plateau region where a contribution of the Pomeron exchange is important for the UA1 data is limited by $|\eta|<2.5$ while the fragmentation rapidity region extends to higher rapidity values. The UA1 data allow exploring the hadroproduction properties in a transition between the central plateau and fragmentation region.

\begin{figure*}[!]
\includegraphics[width =18cm]{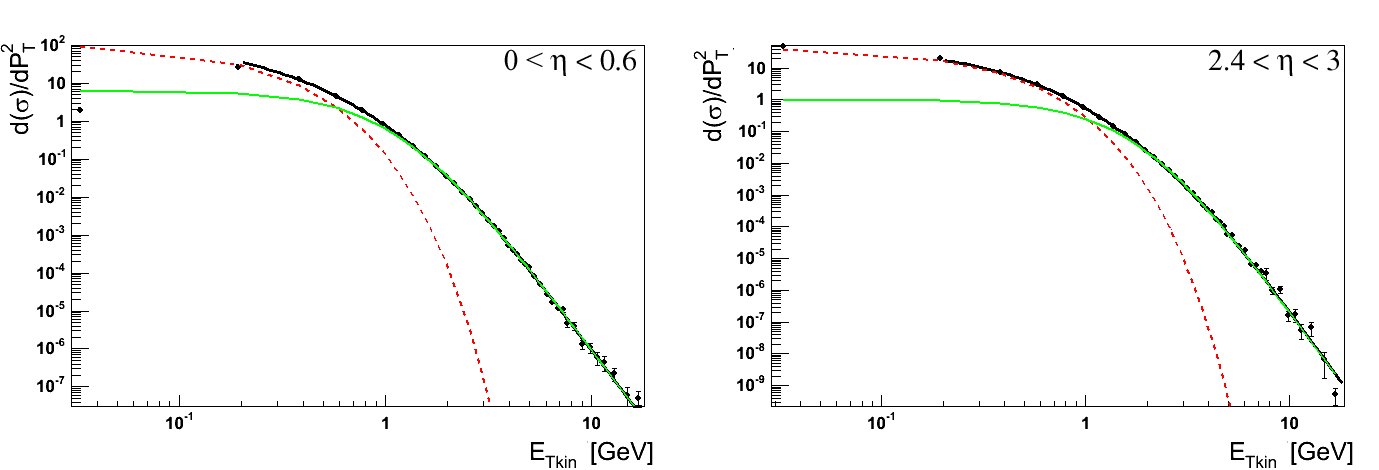}
\caption{\label{fig:02} Charged particle spectrum~\cite{UA1} for different values of pseudorapidity fitted to the function~(\ref{eq:exppl}): the red (dashed) curve shows the exponential term and the green (solid) one stands for the power law term.}
\end{figure*}

\begin{figure}[h]
\includegraphics[width =8cm]{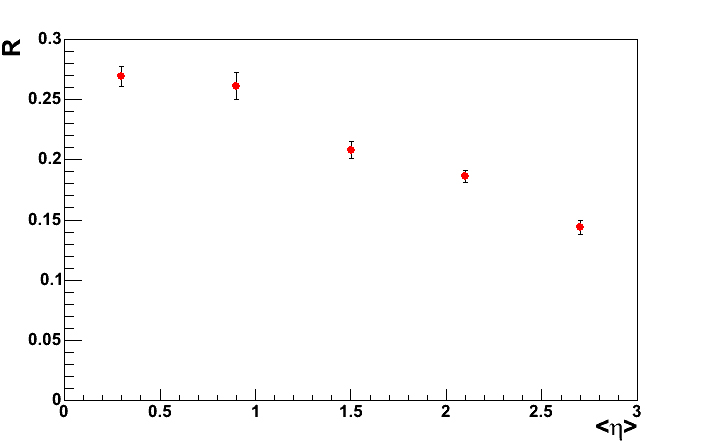}
\caption{\label{fig:03} The relative value of the power law term contribution into the approximation (\ref{eq:exppl}) as function of pseudorapidity.}
\end{figure}

Two examples of charge particle spectra  in different pseudorapidity regions together with the fit to the function (\ref{eq:exppl}) are shown in Figure \ref{fig:02}. It can be seen from this figure that the relative contribution of the exponential part is increasing with the pseudorapidity rising as it is expected from the point of view of the simple model given above. The dependence of the exponential contribution in the approximation (\ref{eq:exppl}) needed to describe the spectrum shape of the produced charged particles as function pseudorapidity is shown in Figure \ref{fig:03}. The functional trend demonstrated in Figure \ref{fig:03} is in accord with the qualitative predictions discussed above. 

In conclusion, a simple naive model for hadroproduction mechanism was proposed to explain a number of recently observed phenomena. Within the framework of this model there are two distinct sources of hadrons produced in particle-particle collisions. One is a radiation of hadrons by the preexisted valence quarks. This source of hadrons is characteristic for colliding massive baryons and is completely missing for colliding gamma quanta.  Another source of hadrons is related to QCD-vacuum fluctuations, described by the Pomeron exchange. The Pomeron interactions give rise to the hadrons distributed according the QCD-like power-law statistical distribution, while the valence quark radiation results in a formation of the exponential Boltzmann-like spectrum of produced particles. This simple model turned out to be successful in describing a number of observed phenomena in hadroproduction. The predicted within this model increase of the relative exponential contribution to the approximation (\ref{eq:exppl}) describing the particle spectra in $pp$-collisions with increasing pseudorapidity was proved on the data previously published by the UA1 Collaboration\cite{UA1}.

The authors thank Professor M.G.Ryskin for fruitful discussion and his help provided during the preparation of this short note.


\end{document}